\documentstyle[preprint,aps]{revtex}
\begin{document}
\title{Dirac and Klein--Gordon particles in complex Coulombic fields; a similarity
transformation.}
\author{Omar Mustafa}
\address{Department of Physics, Eastern Mediterranean University\\
G. Magusa, North Cyprus, Mersin 10 - Turkey\\
email: omar.mustafa@emu.edu.tr}
\maketitle

\begin{abstract}
The observation that the existence of the amazing reality and discreteness
of the spectrum need not necessarily be attributed to the Hermiticity of the
Hamiltonian is reemphasized in the context of the non-Hermitian Dirac and
Klein-Gordon Hamiltonians. Complex Coulombic potentials are considered.
\end{abstract}

In one of the first explicit studies of the non-Hermitian Schr$\ddot{o}$%
dinger Hamiltonians, Caliceti et al [1] have considered the imaginary cubic
oscillator problem in the context of perturbation theory. They have offered
the first rigorous explanation why the spectrum in such a model may be real
and discrete. Only many years later, after being quoted as just a
mathematical curiosity [2] in the literature, the possible physical
relevance of this result reemerged and emphasized [3]. Initiating thereafter
an extensive discussion which resulted in the proposal of the so called $%
{\cal PT}-$symmetric quantum mechanics by Bender and Boettcher [4].

The spiritual wisdom of the new formalism lies in the observation that the
existence of the real spectrum need not necessarily be attributed to the
Hermiticity of the Hamiltonian. This observation has offered a sufficiently
strong motivation for the continued interest in the complex, non-Hermitian,
cubic model which may be understood as a characteristic representation of a
very broad class of the so-called pseudo-Hermitian models with real spectra.

In such non-Hermitian settings, new intensive studies employed , for
example, the idea of the strong coupling expansion [5], the complex \ WKB
[6], Hill determinants and Fourier transformation [7], functional analysis
[8], variational and truncation techniques [9], linear programing [10],
pseudo-perturbation technique [11,12], ..etc (cf [13-15]). However such
studies remain in the context of Schr$\ddot{o}$dinger Hamiltonian and need
to be complemented by the non-Hermitian setting of Dirac and Klein-Gordon
Hamiltonians. Starting, say, with our forthcoming ${\em oversimplified\,}$
generalized complex Coulombic examples.

{\em A priori}, a generalized Dirac - Coulomb equation for a mixed potential
consists of a Lorentz-scalar Coulomb-like and a Lorentz-vector Coulomb
potentials. Whilst the former is added to the mass term of \ the Dirac
equation, the minimal coupling is used, as usual, for the latter. The
ordinary ( Hermitian) Dirac Hamiltonian is exactly solvable in this case (
cf. e.g.,[16,17]). In fact the exact solution to Dirac equation for an
electron in a Coulomb field was first obtained by Darwin [18] and Gordon
[19].

The key idea is that instead of solving Dirac-Coulomb equation directly, one
can solve the second-order Dirac equation [16-22] which is obtained by
multiplying the original equation, from the left, by a differential
operator. The second-order equation is similar to Klein-Gordon equation in a
Coulomb field. The latter reduces to a form nearly identical to that of the
Schr$\ddot{o}$dinger equation and its solution can thus be inferred from the
known non-relativistic solution.

In what follows we recycle the{\em \ modified similarity transformation} (
used by Mustafa et al [17] ) and obtain exact solutions for the
non-Hermitian generalized Dirac and Klein-Gordon Coulomb Hamiltonians.
Although this problem might be seen as {\em oversimplified}, it offers a
benchmark for the {\em yet to be adequately explored} non-Hermitian
relativistic Hamiltonians.

For a mixed scalar and electrostatic complex Coulombic potentials, i.e. $%
m\rightarrow m-iA_{2}/r$ and $V(r)=-iA_{1}/r$, the Dirac Hamiltonian reads (
with the units $\hbar =c=1$)\newline
\begin{equation}
H=\vec{\alpha}\cdot \vec{p}+\beta (m-iA_{2}/r)-iA_{1}/r,
\end{equation}
\newline
where the Dirac matrices $\vec{\alpha}$ and $\beta $ have their usual
meanings. With the {\em similarity transformation} 
\begin{equation}
S=a+i\,b\,\beta \,\vec{\alpha}\cdot \hat{r}\;\;;\;\;S^{-1}=\frac{%
a-i\,b\,\beta \,\vec{\alpha}\cdot \hat{r}}{a^{2}-b^{2}}.\;
\end{equation}
applied to Dirac equation one gets\newline
\begin{equation}
H^{^{\prime }}\Psi ^{^{\prime }}=E\,\Psi ^{^{\prime
}}\;\;;\;\;\,\,H^{^{\prime }}=S\,H\,S^{-1}\;\,,\;\;\Psi ^{^{\prime
}}=S\;\Psi ,
\end{equation}
\newline
\newline
where $\hat{r}$ is the unit vector $\vec{r}/r$ and $a$ and $b$ are constants
to be determined below. For the above central problem, the transformed wave
function is given by\newline
\begin{equation}
\Psi ^{^{\prime }}=\left[ 
\begin{array}{c}
i\,R(r)\,\,\Phi _{jm}^{l} \\ 
Q(r)\,\vec{\sigma}\cdot \hat{r}\,\,\Phi _{jm}^{l}
\end{array}
\right] .
\end{equation}
\newline
In a straightforward manner one obtains, through $E\,\Psi ^{^{\prime
}}=S\,H\,S^{-1}\,\Psi ^{^{\prime }}$, two coupled equations for $R(r)$ ( the
upper component) and $Q(r)$ ( the lower component):\newline
\begin{equation}
\lbrack \partial _{r}+\frac{1}{r}+\frac{K}{r}\,cosh\,\theta +\frac{i\,A_{1}}{%
r}\,sinh\,\theta +E\,sinh\,\theta ]\,R(r)=\xi _{1}(r)\,Q(r),
\end{equation}
\begin{equation}
\lbrack \partial _{r}+\frac{1}{r}-\frac{K}{r}\,cosh\,\theta -\frac{i\,A_{1}}{%
r}\,sinh\,\theta -E\,sinh\,\theta ]\,Q(r)=\xi _{2}(r)\,R(r),
\end{equation}
with 
\begin{equation}
\xi _{1}(r)=m-\frac{i\,A_{2}}{r}+\frac{i\,A_{1}}{r}\,cosh\,\theta +\frac{K}{r%
}\,sinh\,\theta +E\,cosh\,\theta ,
\end{equation}
\begin{equation}
\xi _{2}(r)=m-\frac{i\,A_{2}}{r}-\frac{i\,A_{1}}{r}cosh\,\theta -\frac{K}{r}%
\,sinh\,\theta -E\,cosh\,\theta .
\end{equation}
Where $K=\tilde{\omega}\,(j+1/2)$, $\tilde{\omega}=\mp 1$ for $l=j+\tilde{%
\omega}/2$, $cosh\,\theta =(a^{2}+b^{2})/(a^{2}-b^{2})$, and $sinh\,\theta
=2ab/(a^{2}-b^{2})$.

Incorporating the regular asymptotic behaviour of the radial functions near
the origin; i.e. $R(r)\rightarrow a_{1}r^{\gamma -1}$ and $Q(r)\rightarrow
a_{2}r^{\gamma -1}$ as $r\rightarrow 0$, and neglecting all constant terms
proportional to mass and energy, one obtains\newline
\begin{equation}
\gamma =\sqrt{K^{2}+A_{1}^{2}-A_{2}^{2}}.
\end{equation}
\newline
The negative sign of the square root has to be discarded to avoid divergence
of the wave functions at the origin.

It is obvious that one has the freedom to proceed either with the upper
radial component $R(r)$ or with the lower component $Q(r)$. We shall,
hereinafter, work with the upper component and determine $sinh\,\theta $ and 
$cosh\,\theta $ ( hence the constants $a$ and $b$) by requiring 
\begin{equation}
-i\,A_{2}+i\,A_{1}\,cosh\,\theta +K\,\,sinh\,\theta =0,
\end{equation}
\begin{equation}
K\,cosh\,\theta +i\,A_{1}\,\,sinh\,\theta =\tilde{\omega}\,\gamma ,
\end{equation}
This requirement yields\newline
\begin{equation}
sinh\,\theta =-\,i\,\tilde{\omega}\,\frac{[\,A_{1}\,\gamma -\,|K|\,A_{2}]}{%
[K^{2}+A_{1}^{2}]},\;\;cosh\,\theta =\frac{[\,|K|\,\gamma +A_{1}A_{2}]}{%
[K^{2}+A_{1}^{2}]}.
\end{equation}
Equations (5) and \ (6) would, as a result, imply\newline
\begin{equation}
\lbrack E^{2}-m^{2}]R(r)=\left[ -\partial _{r}^{2}-\frac{2}{r}\partial _{r}+%
\frac{(\gamma ^{2}+\tilde{\omega}\,\gamma )}{r^{2}}-\frac{%
2\,i\,(mA_{2}+A_{1}E)}{r}\right] R(r).
\end{equation}
\newline
With the substitution $R(r)=r^{-1}U(r)$, it reads\newline
\begin{equation}
\lbrack E^{2}-m^{2}]U(r)=\left[ -\partial _{r}^{2}+\frac{(\gamma ^{2}+\tilde{%
\omega}\,\gamma )}{r^{2}}-\frac{2\,i\,(mA_{2}+A_{1}E)}{r}\right] U(r)
\end{equation}
\newline
Evidently this equation is nearly identical to that of the non-Hermitian and 
${\cal PT-}symmetric$ radial Schr$\ddot{o}$dinger - Coulombic one. Of
course, with the irrational angular momentum quantum number $\ell ^{^{\prime
}}=-1/2+\gamma +\tilde{\omega}/2>0$. Its solution can therefore be inferred
from the known non-relativistic ${\cal PT-}symmetric$ Coulomb problem (
c.f.,e.g. Mustafa and Znojil [11] and Znojil and Levai [15] for more details
on this problem). That is \newline
\begin{equation}
\left[ E^{2}-m^{2}\right] ^{1/2}\tilde{n}=[mA_{2}+A_{1}E]\;\;;\;\tilde{n}%
=n_{r}+\ell ^{^{\prime }}+1>0.
\end{equation}
\newline
This in turn implies\newline
\begin{equation}
\frac{E}{m}=\frac{A_{1}A_{2}}{\tilde{n}^{2}-A_{1}^{2}}\pm \left[ \left( 
\frac{A_{1}A_{2}}{\tilde{n}^{2}-A_{1}^{2}}\right) ^{2}+\frac{(\tilde{n}%
^{2}+A_{2}^{2})}{\tilde{n}^{2}-A_{1}^{2}}\right] ^{1/2},
\end{equation}
\newline
with $\tilde{n}=n-j-1/2+\gamma $, where $n_{r}=n-\ell +1$ is the radial
quantum number, $n$ the principle quantum number, and $\ell =j+\tilde{\omega}%
/2$ is the angular momentum quantum number.

In connection with the result in equation (16), several especial cases
should be interesting for they reveal the consequences of the above
complexified non - Hermitian Dirac Hamiltonian:

\begin{description}
\item 
\begin{itemize}
\item  {\em Case 1:} For $A_{2}=0$, the complexified Coulomb energy $%
V(r)=-\,i\,A_{1}/r=-\,i\,Z\,\alpha /r$ ($\alpha \approx 1/137$) represents,
say, the interaction energy of a point nucleus with an imaginary charge $%
iZ\,e$ and a particle of charge $-e$. In this case $\gamma =\sqrt{%
(j+1/2)^{2}+(Z\alpha )^{2}}$, and 
\begin{equation}
\frac{E}{m}=+\;\left[ 1-\frac{(Z\alpha )^{2}}{(n-j-1/2+\gamma )^{2}}\right]
^{-1/2},
\end{equation}
where the negative sign is excluded because negative energies would not
fulfill equation (15). For a vanishing potential ( $Z=0$ ) the energy
eigenvalue is $m$. Obviously, unlike the ordinary ( Hermitian){\em \
Sommerfeld} fine structure formula, equation (17) suggests that a continuous
increase of the coupling strength $Z\alpha $ from zero {\em pushes up }the
electron states into the positive energy continuum, avoiding herby the
energy gap. Nevertheless, for states with $n=j+1/2$ one obtains 
\begin{equation}
\frac{E}{m}=+\;\sqrt{1+\frac{(Z\alpha )^{2}}{n^{2}}},
\end{equation}
The ratio $E/m$ in (18) is plotted in figure 1 for $n=1,2,3,..,10,20,..50$.
It is evident that as $n\rightarrow \infty $ the ratio $E/m\rightarrow 1$.

\item  {\em Case 2 : }For $A_{1}=0$, $\gamma =\sqrt{K^{2}-A_{2}{}^{2}}$ and
equation (16) reads 
\begin{equation}
\frac{E}{m}=\pm \;\left[ 1+\frac{A_{2}{}^{2}}{(n-j-1/2+\gamma )^{2}}\right]
^{1/2},
\end{equation}

In this case both signs are admissible and thus two branches of solutions
exist, but not in the energy gap. The solutions of positive and negative
energies exhibit identical behaviour, which reflects the fact that scalar
interactions do not distinguish between positive and negative charges.
Moreover, states with negative energies are {\em pulled down} to dive into
the negative energy continuum, while states with positive energies are {\em %
pushed up} to dive into the positive energy continuum. Yet, the {\em flown
away states }phenomenon reemerges and for $A_{2}=|K|$ states with $n=j+1/2$ 
{\em fly away} and disappear from the spectrum. Of course one should worry
about the critical values of the coupling (i.e., $A_{2,\,crit}=|K|$) where
imaginary energies would be manifested.

\item  {\em Case 3 :} For $A_{1}=A_{2}=A,$ $\gamma =|K|$ and

\begin{equation}
\frac{E}{m}=\frac{A^{2}}{\tilde{n}^{2}-A^{2}}\pm \frac{\tilde{n}^{2}}{\tilde{%
n}^{2}-A^{2}},
\end{equation}

Obviously, the negative sign must be discarded for it implies $E=-m$ and
thus contradicts equation (15). Hence, equation (20) reduces to 
\begin{equation}
\frac{E}{m}=1+\frac{2\,A^{2}}{n^{2}-A^{2}}.
\end{equation}

Part of this spectrum (i.e. for the principle quantum number $n=1,2,3,..,6$
) is plotted in figure 2. As the coupling strength $A$ increases from zero
to $n$, the electron states are {\em pushed up }from $E=m$ into the positive
energy continuum avoiding the energy gap between $-m$ to $m$. However, all
states with $n=A$ {\em fly away }and disappear from the spectrum.
Nevertheless, as $A$ increases from $n$ and at $A\rightarrow $ $\infty $ all
energy states ${\em cluster}$ just below $E=-\,m$.

\item  {\em Case 4 : }If we replace $(\gamma ^{2}+\tilde{\omega}\,\gamma )$
with $\tilde{\ell}\,(\tilde{\ell}+1)$, where $\tilde{\ell}=-1/2+\sqrt{(\ell
+1/2)^{2}+A_{1}^{2}-A_{2}^{2}}$ , equation (14) reduces to Klein-Gordon [23]
with complex Coulomb-like Lorentz scalar and Lorentz vector potentials, $%
S(r)=-i\,A_{2}/r$ and $V(r)=-i\,A_{1}/r$, respectively. That is 
\begin{equation}
\lbrack E^{2}-m^{2}]U(r)=\left[ -\partial _{r}^{2}+\frac{\tilde{\ell}\,(%
\tilde{\ell}+1)}{r^{2}}-\frac{2\,i\,(mA_{2}+A_{1}E)}{r}\right] U(r).
\end{equation}
\newline

Which when compared with the non-Hermitian ${\cal PT-}symmetric$ Schr$\ddot{o%
}$dinger-Coulomb equation implies that 
\begin{equation}
\left[ E^{2}-m^{2}\right] ^{1/2}\tilde{N}=[mA_{2}+A_{1}E]\;\;;\;\tilde{N}%
=n_{r}+\tilde{\ell}+1>0,
\end{equation}

and 
\begin{equation}
\frac{E}{m}=\frac{A_{1}A_{2}}{\tilde{N}^{2}-A_{1}^{2}}\pm \left[ \left( 
\frac{A_{1}A_{2}}{\tilde{N}^{2}-A_{1}^{2}}\right) ^{2}+\frac{(\tilde{N}%
^{2}+A_{2}^{2})}{\tilde{N}^{2}-A_{1}^{2}}\right] ^{1/2}.
\end{equation}

This in turn, following similar analysis as above, yields 
\begin{equation}
\frac{E}{m}=+\,\left[ 1-\frac{A_{1}^{2}}{\tilde{N}^{2}}\right] ^{-1/2}\;;\;%
\tilde{N}=n-\ell -1/2+\sqrt{(\ell +1/2)^{2}+A_{1}^{2}},
\end{equation}

for $A_{2}=0$ and $A_{1}\neq 0,$%
\begin{equation}
\frac{E}{m}=\pm \,\left[ 1+\frac{A_{2}^{2}}{\tilde{N}^{2}}\right] ^{1/2}\;;\;%
\tilde{N}=n-\ell -1/2+\sqrt{(\ell +1/2)^{2}-A_{2}^{2}},
\end{equation}

for $A_{1}=0$ and $A_{2}\neq 0,$ and 
\begin{equation}
\frac{E}{m}=+\,\left[ 1+\frac{2\,A^{2}}{n^{2}-A^{2}}\right] \;;\;\tilde{N}=n,
\end{equation}
for $A_{1}=A_{2}=A$. Clearly, spin-0 states follow similar scenarios as
those for spin-1/2 states ( i.e., e.g., {\em flown away states, pushed up
into the \ positive continuum and/or pulled down into the negative continuum
..}etc.).

To summarize, we have used a similarity transformation to extract exact
energies for Dirac - particle in the generalized complex Coulomb potential.
Whithin such non-Hermitian settings we have also obtained exact energies for
Klein-Gordon particle.
\end{itemize}
\end{description}

\newpage

\begin{center}
{\bf {\huge Figures Captions}}
\end{center}

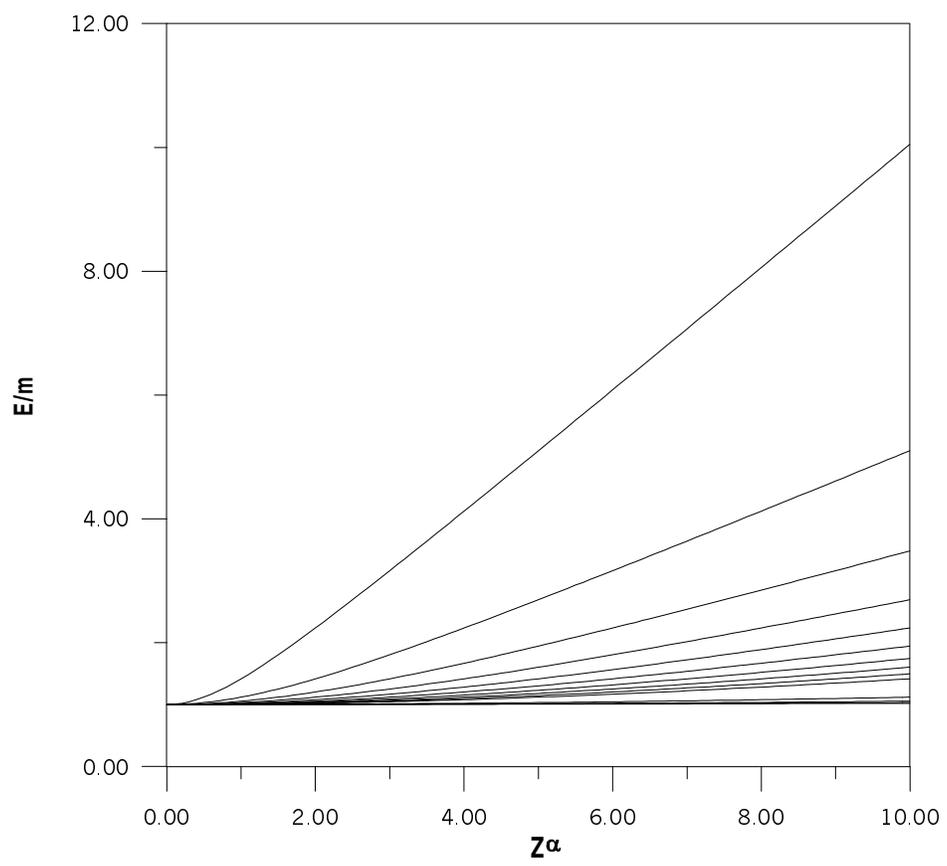
{\bf Figure 1}. The ratio $E/m$ of (18) at different vales of $Z\alpha $ for
the states (from top to bottom) with the principle quantum number $%
n=1,2,3,...,10,20,30,40,$and $50$.

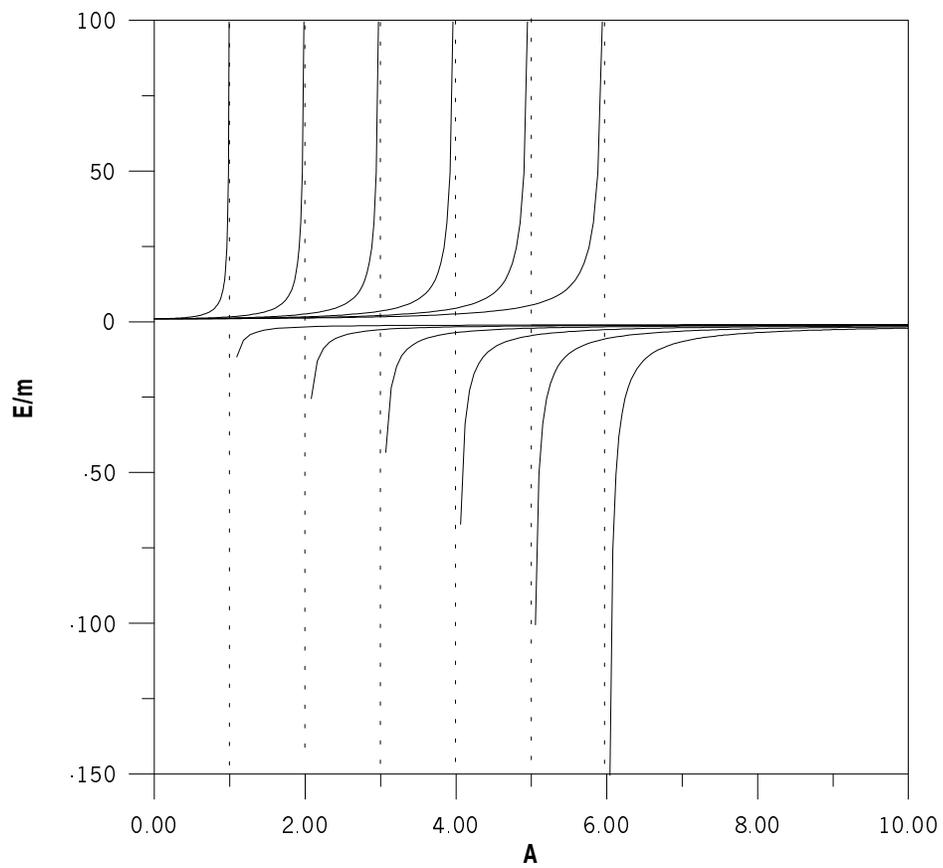
{\bf Figure 2}. Part of the spectrum $E/m$ of (21) at different values of
the coupling $A$ and for states with $n=1,2,3,..,6$.

\end{document}